\documentclass[preprint,preprintnumbers,amsmath,amssymb]{revtex4}
\usepackage{graphicx}
\usepackage{adjustbox}
\usepackage{epstopdf}    
\usepackage{subcaption} 
\usepackage{dcolumn}
\usepackage{bm}
\usepackage{hyperref}
\hypersetup{
	colorlinks=true,
	linkcolor=black,
	citecolor=blue,
	pdfborder={0 0 0},
	backref=true
}

\begin{document}
	
\title{Decoherence of ergotropy of a relativistic battery as a probe of motion-selected Unruh thermality in Minkowski spacetime}
\author{Tao-Feng Gan}
\author{Cheng-Tai Wu}
\author{Yin-Zhong Wu}
\email{yzwu@mail.usts.edu.cn}
\author{Xiang Hao}
\altaffiliation{Corresponding author}
\email{xhao@mail.usts.edu.cn}
\affiliation{School of Physical Science and Technology, Suzhou University of Science and Technology, Suzhou, Jiangsu 215009, People's Republic of China}
\affiliation{Pacific Institute of Theoretical Physics, Department of Physics and Astronomy,
\\University of British Columbia, 6224 Agriculture Rd., Vancouver B.C., Canada V6T 1Z1.}
	
\begin{abstract}

We put forward a physical model of a relativistic Unruh-DeWitt battery moving along a general accelerated trajectory, including a linear motion and a circular motion with a reflecting boundary. The maximal amount of quantum work extraction, defined as the ergotropy, serves as a witness to Unruh effect modified by motion trajectories. The asymptotic behavior of ergotropy in a linear motion can demonstrate the Unruh thermality with respect to the Kubo-Martin-Schwinger condition, which is independent of the presence of a boundary. It is found out that for a very low Unruh temperature, linear motion yields a high amount of ergotropy, while for a high temperature, circular motion becomes optimal for estimating the Unruh effect. For a specific acceleration, the ergotropy is the same for two different trajectories. The behavior demonstrates that, in a certain condition, one can simulate the work extraction of the accelerated battery in a linear motion by exploring the battery in circular motion.  The observed ergotropy for the thermality is closely related to quantum coherence affected by spacetime vacuum fluctuations. In the presence of a reflecting boundary plane, we study different ways for the dynamics of the ergotropy. When the battery moves near the boundary, the prominent oscillation of the ergotropy will happen and exhibit some large instantaneous peaks. The interesting phenomenon results from the protection of quantum coherence for the accelerated battery in the vicinity of the boundary. Far away from the boundary, the oscillation behavior can be suppressed and the ergotropy rapidly arrives at a steady value. By contrast, the circular motion contributes to prolonging the oscillation evolution. The asymptotic amount of the ergotropy can rise progressively up to a saturation value with increasing the distance from the circular trajectory to the boundary plane. From the perspective of energy transfer, optimized quantum work extraction for the accelerated battery moving along a selected motion with a boundary plane is advantageous for probing Unruh thermality.
		
\end{abstract}
	
\maketitle
	
\section{Introduction}
As one fundamental phenomenon of quantum field theory in curved spacetimes, Unruh found that a uniformly accelerated observer can detect the Minkowski vacuum as a thermal bath of a Planckian spectrum with respect to a temperature proportional to the acceleration \cite{unruh1976notes,hawking1975particle}. The Unruh effect provides deep insights into the quantum nature of black hole evaporation and cosmological horizon \cite{gibbons1977cosmological,liu2016quantum}. The Unruh effect originating from vacuum fluctuations can be investigated through the response spectrum of the Unruh-DeWitt (UDW) detector moving along a certain trajectory \cite{fulling1973nonuniqueness,davies1975scalar}. The UDW detector has been proved to be a useful quantum probe for characterizing vacuum fluctuations by means of open quantum system thermodynamics \cite{crispino2008unruh}. However, the estimation of Unruh temperature remains challenging in current experimental setups due to the extremely high acceleration required for experimental observations \cite{bell1983electrons,cardi2025measurement}. According to the thermalization theorem, an accelerated detector experiences the loss of unitarity, guaranteed by the Kubo-Martin-Schwinger (KMS) condition which does not imply a Planckian spectrum in massive scalar background or a high dimension spactime\cite{takagi1986vacuum,sriramkumar2003interpolating,arrechea2021inversion}. Motivated by these open questions, many researchers have focused on the interplay between relativistic effects and quantum information science, paving a new way for Unruh thermality \cite{mann2012relativistic}. Relevant studies have explored the thermal properties of the Unruh effect from various perspectives, including quantum nonlocality \cite{pozas2015harvesting,feng2018bell,li2025quantifying}, geometric phase \cite{hu2012geometric,zhao2022geometric}, and quantum parameter estimation \cite{huang2019decoherence,liu2021does,patterson2023fisher}. These methods allow for a comprehensive examination of the connection between quantum correlations and Unruh thermal effect. Nevertheless, many existing conclusions are largely confined to a linear acceleration trajectory. Its validity for other large acceleration scenarios have rarely been explored. Moreover, the effect of a reflecting boundary on the measurement of Unruh effect is also valuable. Our work endeavors to provide an alternative framework for a better performance in detecting Unruh thermality when employing the optimal motion trajectory in light of quantum energy transfer.

Quantum battery is referred to as a quantum device capable of storing energy from external chargers by utilizing quantum resources \cite{alicki2013entanglement,campaioli2017enhancing,le2018spin,ferraro2018high,zhang2019powerful,garcia2020fluctuations,farina2019charger,ghosh2021fast,cruz2022quantum,barra2022quantum,rossini2020quantum,chen2025quantum,zhang2025coherent,Liu2025SciChina,Tian2025JHEP}. For a quantum battery in a non-passive state, quantum work can be extracted through cyclic unitary operations. Conversely, if the battery is in a passive state, no work can be obtained through such operations. The maximal amount of extractable work, defined as ergotropy, is a crucial metric for charging performance of a quantum battery. For a $d$-dimensional quantum battery~\cite{alicki2013entanglement}, the internal Hamiltonian $H_0=\sum_{j=1}^{d}\epsilon_j |j\rangle \langle j|$ has non-degenerate energy levels. To extract the work from the battery, one need to apply a cyclic unitary operation by an external charging $H_c(t)$ where the charging process is turned on at time $t=0$ and off at $t=\tau$. During the charging period $\tau$, the work extraction is dependent on the evolved state $\rho(\tau)$ of the battery which is determined by $\rho(\tau)=U(\tau)\rho(0)U^{\dag}(\tau)$, and $U(\tau)=\mathcal{T}\exp \{ -i \int_0^{\tau} \mathrm{d}s [H_0+H_c(s)] \}$ is the evolution operator in the time ordering operator $\mathcal{T}$. The average work extracted from the battery is defined as $W=\mathrm{Tr}\{ H_0[\rho(0)-\rho(\tau)]\} $ and the maximal value of $W$ over the set of any possible unitary operation is called the ergotropy. 

In some recent studies, the UDW battery models in a flat spacetime were adopted to explore the influences of acceleration‑induced thermal effects on charging performance~\cite{Liu2025SciChina} with a reflecting boundary condition. The relativistic battery was also extended to curved spacetime background in order to investigate the effect of black hole geometry on the dynamics of the ergotropy~\cite{Tian2025JHEP}. In this paper, we propose an accelerated UDW battery and explore quantum extractable work in the condition of different motion trajectories without or with a reflecting boudary. The extractable work of the UDW battery serves as a witness to the Unruh thermal effect. We view the UDW battery as the open quantum system immersed in an environment of fluctuating vacuum scalar fields. It is known that the vacuum fluctuations in Minkowski spacetime can be modified by the acceleration profiles of the detector. Furthermore, the presence of a reflecting boundary also impacts on the estimation of Unruh thermality. By exploring the asymptotical and dynamical behavior of quantum work extraction, we expect to reveal the global and position-dependent characteristics of Unruh effect under the circumstances of different acceleration trajectories.

Our motivation for utilizing quantum work extraction to explore the Unruh thermal effect stems primarily from the following reasons. Firstly, quantum work extraction serves as an operational measure of energy transfer between a battery and its environment. The spacetime properties and relativistic motions may be encoded into the dynamical evolution of extractable work which can act as a sensitive probe for the thermal nature of Unruh effect. Secondly, the behavior of quantum work extraction is intricately linked to the quantum resources of relativistic quantum batteries. Existing studies indicate that the fundamental reason for the Unruh effect exhibiting thermal characteristics lies in its capture of quantum correlations near the Rindler horizon, as well as quantum fluctuations of the Minkowski vacuum field. Within this context, observing high stable values of quantum work extraction with respect to different trajectories will facilitate the analysis of Minkowski vacuum fluctuations. Therefore, we employ quantum work extraction of the accelerated UDW battery to address several fundamental and previously neglected questions. The first one concerns which acceleration scenarios can enhance quantum work extraction to improve the estimation of vacuum fluctuations. The second question concerns how the boundary influences the detection efficiency for different motion trajectories via a relativistic quantum battery. Through a systematic exploration of these questions, we intend to gain deeper insight into relativistic quantum batteries and the implications of such batteries for the detection of Unruh thermality in accelerated reference frames.

We employ natural units $G=c=\hbar=k_{B}=1$ throughout the paper. The outline of the paper is as follows. In section 2, we construct a physical model of a relativistic quantum battery which moves along a general motion trajectory in Minkowski spacetime. By means of open quantum system approach, we derive a quantum master equation to explore the relationship between trajectory-dependent response spectrum and quantum work extraction. In section 3, we investigate the behavior of the ergotropy for the relativistic battery in different acceleration scenarios including uniformly linear accelerated and circular motion. The asymptotical properties of the ergotropy demonstrate the global characteristic of Unruh thermality modified by different trajectories. In section 4, the effects of a reflecting boundary for different motions on the evolution of the ergotropy are studied in detail. Finally, in section 5, we present the conclusions and discussions.

\section{The UDW battery model}
We employ a relativistic quantum system driven by an external field in Minkowski spacetime. The probe system consists of a two-level detector undergoing possible motions along some adjustable trajectories. The accelerated battery can be viewed as an open quantum system coupled to a scalar vacuum field in Minkowski spacetime. The battery is generated by the Hamiltonian of $H_0=\frac {\omega_0}2 \sigma_{3}$ where $\omega_0$ denotes the transition frequency between an excited state $|e\rangle$ and a ground state $|g\rangle$. The operator $\sigma_{j=1,2,3}$ represents the $j$-component of the Pauli operator. To realize a charging process, we need apply a field to drive the battery~\cite{Liu2025SciChina,Tian2025JHEP}. The external charging Hamiltonian $H_c(t)=\mu(t)\frac {\omega}2 \sum_{j=1}^{3} n_j \sigma_j$ where the unit direction of the driving field is represented by $\mathbf{n}=(n_1,n_2,n_3),\; |\mathbf{n}|=1$ and $\omega$ denotes the strength. Here, $\mu(t)=1(0<t<\tau)$ is is the switching function for the charging process. In this protocol, a relativistic UDW battery is regarded as an open system which is coupled to a fluctuating vacuum field in Minkowski spacetime. 

The total Hamiltonian of the UDW battery along a general motion trajectory $x(\tau)$ in curved spacetime can be expressed as
\begin{equation}
	\label{eq:(1)}
	H=H_0+H_c+H_{\phi}+H_{I}.
\end{equation}		
The Hamiltonian $H_{\phi}$ composed of the scalar field operator $\Phi (x(\tau ))$ satisfies the standard Klein-Gordon equation in Minkowski spacetime. The interaction between the battery and vacuum field $H_I=\mu(t)\lambda(\cos \delta \sigma_1+\sin \delta \sigma_2)\Phi \left( {x(\tau )} \right)$ is assumed to be linear~\cite{Liu2025SciChina,Tian2025JHEP}. During the charging proper time $\tau$, the switching function $\mu(t)$ is used to control the interplay between the battery and external fields. The parameter $\lambda$ represents the weak coupling strength and $\delta$ is used to evaluate the weight of the battery-field coupling which captures the decoherence process.

In order to investigate the dynamics of quantum battery, we consider a unitary rotation in the spin space, represented as $R(\theta)=e^{i\frac {\theta}{2}\sigma_3}$ where $\theta$ is carefully selected to ensure the combined Hamiltonian can be projected only along the $x$-axis. By using the rotation, we can obtain the following, $\tilde{H}_0=RH_0R^{\dag}=H_0$ and $\tilde{H}_c=RH_cR^{\dag}=\frac {\omega}2[(n_1 \cos\theta + n_2 \sin \theta)\sigma_1+(n_2 \cos\theta-n_1 \sin\theta)\sigma_2+n_3\sigma_3]$. Here, the adjustable angle $\theta$ and external field satisfy the conditions of $\cos \theta=\frac {n_1}{\sqrt{1-(n_3)^2}}, \sin \theta=\frac {n_2}{\sqrt{1-(n_3)^2}}$, and $n_3=-\frac {\omega_0}{\omega}$. Under the above conditions, the transformed Hamiltonian of the battery-charger combined system is expressed as 
\begin{equation}
\label{eq:(2)}
H_b=R(H_0+H_c)R^{\dag}=\frac {\Omega}2 \sigma_1,
\end{equation}
where $\Omega=\omega(\sqrt{1-(n_3)^2})$ denotes the effective coupling strength. In the rotating frame, $\tilde{H}_{I} = R H_I R^{\dag}$ represents the rotated Hamiltonian describing the interaction between the battery and fluctuating field. In the rotation frame, the part of $\tilde{H}_I=\lambda[\cos (\delta-\theta)\sigma_1+\sin (\delta-\theta)\sigma_2]\Phi \left( {x(\tau )} \right)$. When $\delta=\theta$, the rotated interaction Hamiltonian can be written as 
\begin{equation}
\label{eq:(3)}
\tilde{H}_I=\lambda \sigma_1 \Phi \left( {x(\tau )} \right). 
\end{equation}
In the condition of weak couplings, i. e., $\lambda \ll \Omega$, the UDW battery is charged by both the coherent driving field and the scalar field in curved spacetimes.

As an initial state, the battery $\tau=0$ in the ground state $\left| g \right\rangle $ indicates a depleted condition. The ground state $\rho(0)=\frac {I+\sum_j w_j(0)\sigma_j}{2}$, where the components of the Bloch vector are expressed by $w_1(0)=w_2(0)=0, w_3(0)=-1$. In the rotating frame, $\tilde{\rho}(0)=R\rho(0)R^{\dag}=\frac {I+\sum_j r_j(0)\sigma_j}{2}$ is a state after performing the rotation $R$ on the ground state. The corresponding Bloch vector of $\tilde{\rho}(0)$ can be obtained by $r_1(0)=w_1(0)\cos \theta+w_2(0)\sin \theta$, $r_2(0)=w_2(0)\cos \theta-w_1(0)\sin \theta$ and $r_3(0)=w_3(0)$. In the condition of weak couplings, the initial state of the combined system can be approximated as $\tilde{\rho}_{tot}(0)=\tilde{\rho}(0)\otimes |0\rangle \langle 0|$, where $|0\rangle$ denotes the vacuum state of a massless scalar field in curved spacetime. For the whole system, the dynamics of the total state can be governed by $\frac{\partial \tilde{\rho}_{tot}(\tau)}{\partial \tau}=-i[\tilde{H}_I(\tau), \tilde{\rho}_{tot}(\tau)]$ where $\tilde{H}_I(\tau)$ denotes the interaction Hamiltonian of the battery-field combined system in the interaction picture. From the viewpoint of open quantum systems, the scalar vacuum field is considered as the environment. Accordingly, the reduced density matrix of the battery $\tilde{\rho}(\tau)=Tr_{f}[\tilde{\rho}_{tot}(\tau)]$ is obtained by tracing over the environmental degrees of freedom.

As a probe system, a specific motion trajectory can induce some modifications to the vacuum quantum fluctuations. Therefore, the optimal selection of feasible trajectory plays a role in the charging performance. We assume that the two-level battery moves along a specific trajectory of $x(\tau )$ parameterized by the proper time $\tau$ which is also viewed as the charging time. The charging protocol can be influenced by both the driving field and acceleration motion in Minkowski spacetime. In the condition of weak interactions, the dynamics of the battery is governed by the quantum master equation \cite{benatti2004entanglement} in the Gorini-Kossakowski-Sudarshan-Lindblad(GKSL) form,
\begin{align}
\label{eq:(4)}
	\frac {\partial}{\partial \tau} \tilde{\rho}(\tau)&\;=\;-i[H_{b}^{(\mathrm{eff})}, \tilde{\rho}(\tau)]+\frac 12\sum_{i,j=1}^{3}a_{ij}\mathcal{D}_{ij}[\tilde{\rho}(\tau)], \\
	a_{ij}&\;=\;A\delta_{ij}-iB\varepsilon_{ijk}\delta_{k1}+C\delta_{i1}\delta_{j1}, \nonumber \\
	A&\;=\; \frac {\lambda^2}{2}[\mathcal{G}(\Omega)+\mathcal{G}(-\Omega)],\;B= \frac {\lambda^2}{2}[\mathcal{G}(\Omega)-\mathcal{G}(-\Omega)],\;C=\lambda^2 \mathcal{G}(0)-A,\nonumber
\end{align}
where the dissipator $\mathcal{D}_{ij}(\tilde{\rho})=2\sigma_j\tilde{\rho}\sigma_i-\sigma_i\sigma_j\tilde{\rho}-\tilde{\rho}\sigma_i\sigma_j$ arises from the dissipation and decoherence induced by the environment. By introducing the Wightman function of scalar field $ G^{+}(x-x')=\langle 0|\Phi \big( x(\tau) \big)\Phi \big( x'(\tau^{'}) \big)|0\rangle=\frac {1}{4\pi^2[|\vec{x}-\vec{x}'|^2-(t-t'-i\epsilon)]}$, we can derive its Fourier transform
\begin{equation}
	\label{eq:(5)}
	\mathcal{G}(\Omega)=\int_{-\infty}^{\infty} \mathrm{d}\Delta \tau \cdot e^{i\Omega\Delta \tau} G^{+}(\Delta \tau).
\end{equation}
The Hilbert transform of the Wightman function is given by $\mathcal{K}(\Omega)=\frac {\mathcal{P}}{\pi i}\int_{-\infty}^{\infty} \mathrm{d}\omega\frac {\mathcal{G}(\omega)}{\omega-\Omega}$ where $\Delta \tau=\tau-\tau'$ and $\mathcal{P}$ denotes the principle value. The effective Hamiltonian is given by $H_{b}^{(\mathrm{eff})}=\frac {1}{2}\Omega^{'}(\sigma^{+}+\sigma^{-})$ with $\Omega^{'}=\Omega+i\lambda^2[\mathcal{K}(-\Omega)-\mathcal{K}(\Omega)]$ representing the effective coupling. The interaction with external scalar field would have an effect on the Lamb shift. In the case of weak couplings, $\lambda \ll \Omega$, we can neglect the Lamb shift in the following analysis.

 Utilizing the Lindblad master equation, we can obtain the dynamics of the battery via the Bloch equation,
\begin{equation}
	\label{eq:(6)}
	\frac{d}{{d\tau }}\mathbf{r}(\tau ) =  - 2\bm{\mathcal{H}} \cdot \mathbf{r}(\tau ) + \bm{\chi}(\tau),
\end{equation}
where $\bm{\mathcal{H}}$ is a decaying matrix in the form of
\begin{equation}
	\label{eq:(7)}
	\bm{\mathcal{H}} = \left( {\begin{array}{*{20}{c}}
			{2A}&0&0\\
			0&{2A + C}&{\Omega /2}\\
			0&{ - \Omega /2}&{2A + C}
	\end{array}} \right).
\end{equation}
Here the inhomogeneous vector $\bm{\chi}(\tau )= ( - 4B,0,0)^{\rm T}$. We use a quantum channel to characterize the evolution of the battery
\begin{equation}
	\label{eq:(8)}
	\mathbf{r}(\tau ) = \bm{\Gamma}(\tau) \cdot\mathbf{r}(0) + \mathbf{\Lambda}(\tau ),
\end{equation}
where the mapped matrix of the quantum channel $\bm{\Gamma} (\tau )=\exp (-2\bm{\mathcal{H}}\tau)$ is expressed as
\begin{equation}
	\label{eq:(9)}
	\bm{\Gamma} (\tau ) =  \begin{pmatrix}
		 {\begin{array}{*{20}{c}}
				{{e^{ - 4A\tau }}}&0&0\\
				0&{{e^{ - 2(2A + C)\tau }}\cos \Omega \tau }&{ - {e^{ - 2(2A + C)\tau }}\sin \Omega \tau }\\
				0&{{e^{ - 2(2A + C)\tau }}\sin \Omega \tau }&{{e^{ - 2(2A + C)\tau }}\cos \Omega \tau }
		\end{array}}
	\end{pmatrix}.
\end{equation}
and $\bm{\Lambda} (\tau ) = \frac{1}{2}[I - \bm{\Gamma} (\tau )]{\bm{\mathcal{H}}^{ - 1}}\cdot \bm{\chi}$ is the mapped vector. The Unruh channel can be established by using the mapped matrix and vector.

When the initial state $r_1(0)=r_2(0)=0,r_3(0)=-1$, the vector can be expressed as
\begin{equation}
	\label{eq:(10)}
	\mathbf{r}(\tau ) = \left( \begin{array}{l}
		\tilde{\gamma} ({e^{ - 4A\tau }} - 1)\\
		{e^{ - 2(2A + C)\tau }}\sin \Omega \tau \\
		- {e^{ - 2(2A + C)\tau }}\cos \Omega \tau
	\end{array} \right).
\end{equation}
The ratio $\tilde{\gamma}=\frac {B}{A}$ is determined by the field Wightman function which is dependent on the vacuum fluctuation induced by the accelerated motion in Minkowski spacetime. In the asymptotical condition of $\tau \rightarrow \infty$, the battery evolves into a steady state as a result of thermalization, where $\mathbf{r}^{(s)}=-\tilde{\gamma}(1,0,0)^{\mathrm{T}}$ is closely connected with the Unruh thermality. Meanwhile, the simultaneous characteristic of the evolution of the battery is determined by both the trajectory and driving field.

In the condition of the rotating transformation, the maximal amount of quantum work extraction over all unitary transformations $\{ \tilde{U}(\tau)  \}$, i.e., the ergotropy, can be written as $\mathcal{W}(\tau)=\mathrm{Tr}[\tilde{H}_0\tilde{\rho}(\tau)]-\min_{\{ \tilde{U} \}}\mathrm{Tr}[\tilde{U}\tilde{\rho}(\tau)\tilde{U}^{\dag}\tilde{H}_0]$.
Note that the ergotropy remains unchanged because $\mathrm{Tr}[H_0\rho(0)]=\mathrm{Tr}[\tilde{H}_0\tilde{\rho}(0)]$ and $\mathrm{Tr}[U\rho(\tau)U^{\dag}H_0]=\mathrm{Tr}[\tilde{U}\tilde{\rho}(\tau)\tilde{U}^{\dag}\tilde{H}_0]$. The minimal unitary transformation $\tilde{U}_{\sigma}$ is optimized to transform the battery into a passive state where no work can be extracted. The passive state satisfies that  $\tilde{U}_{\sigma}\tilde{\rho}(\tau)\tilde{U}^{\dag}_{\sigma}=\sum_{j}\varrho_j|\epsilon_j\rangle  \langle \epsilon_j |$ where $|\epsilon_j\rangle $ is the energy level state of $\tilde{H}_0$ with the corresponding energy $\epsilon_j$ in the increasing order and the eigenvalues $\varrho_j$ of $\tilde{\rho}(\tau)$ are arranged in the decreasing order. Therefore, the simplified expression of the ergotropy~\cite{alicki2013entanglement} is given in the form of
$\mathcal{W}(\tau)=\sum_{j,k}\varrho_j \epsilon_k (|\langle \varrho_j|\epsilon_k\rangle|^2-\delta_{jk}).$
The larger maximal amount of extractable work, the better charging performance of quantum battery.

To express the ergotropy, we should firstly calculate the eigenvalues $\varrho_{1,2}=\textstyle{{1 \pm \left| {\mathbf{r}(\tau )} \right|} \over 2}$. The optimal unitary operation is expressed as $\tilde{U}_\sigma =\sum\nolimits_{j=1,2} {\left|  \epsilon_j\right\rangle \left\langle \varrho_j \right|}$ where $\left|  \varrho_j\right\rangle$ is the eigenvector of the rotated density matrix of the battery state. In the space of $\{|\epsilon_1\rangle=|g\rangle ,|\epsilon_2\rangle=|e\rangle\}$, $|\varrho_{1}\rangle=\sqrt{\frac {r+r_3}{2r}}|g\rangle+\frac {r_1+ir_2}{\sqrt{2r(r+r_3)}}|e\rangle$ and $|\varrho_{2}\rangle=\frac {r_1-ir_2}{\sqrt{2r(r+r_3)}}|g\rangle-\sqrt{\frac {r+r_3}{2r}}|e\rangle$ are obtained. The ergotropy can be achieved by the optimal unitary transformation,
\begin{equation}
	\label{eq:(11)}
	\tilde{U}_{\sigma}=\left( \begin{array}{cc} \sqrt{\dfrac {r+r_3}{2r}} & \dfrac {r_1-ir_2}{\sqrt{2r(r+r_3)}} \\ \dfrac {r_1+ir_2}{\sqrt{2r(r+r_3)}} & - \sqrt{\dfrac {r+r_3}{2r}}  \end{array}  \right),
\end{equation}
where $r=\left| {\mathbf{r}(\tau )} \right|$ represents the norm of the Bloch vector. The value of the ergotropy $\mathcal{W}$ is determined by the internal energy $E(\tau)=\mathrm{Tr}[\tilde{H}_0\tilde{\rho}(\tau)]=\frac {\omega_0(1+r_3)}2$ and the part of $\mathrm{Tr}[\tilde{U}_{\sigma}\tilde{\rho}(\tau)\tilde{U}_{\sigma}^{\dag}\tilde{H}_0]=\mathrm{Tr}[\tilde{\rho}_{\sigma}\tilde{H}_0]=\sum_{j=1,2}\varrho_j \epsilon_j= \frac {\omega_0(1-r)}2$. Therefore, the ergotropy is expressed as $\mathcal{W}=\frac {\omega_0(r+r_3)}2$. For convenience, we can define the scaled ergotropy as $\xi(\tau)=\frac {\mathcal{W}}{\omega_0}$,
\begin{equation}
	\label{eq:(12)}
	\xi(\tau)=\frac {1}{2}[r(\tau)+r_3(\tau)].
\end{equation}
The time-dependent Bloch vector $\mathbf{r}(\tau)$ is obtained by the dynamics of the rotated density matrix of the charged battery moving in Minkowski spacetime.
In the asymptotic condition of $\tau \rightarrow \infty$, we can obtain the steady-state ergotropy after Unruh thermalization,
\begin{equation}
	\label{eq:(13)}
	{\xi^{(s)}} = \frac{\tilde{\gamma} }{2}.
\end{equation}
It is noticed that the asymptotic ergotropy is determined by the Unruh thermality and driving field. In addition, the thermal nature of Unruh effects arises from the vacuum fluctuations, which can be explained in terms of Fock-space states where quantum coherence exists across the causal horizon of the accelerated observer. This evidence motivates us to explore the relationship of acceleration-affected quantum coherence and quantum work extraction. According to the result of \cite{baumgratz2014quantifying}, quantum coherence for a two-level system can be expressed by $C(\rho)=\sum_{j\neq l}\rho_{jl}=\sqrt{r_1^2+r_2^2}$ where $r_j$ is the component of the Bloch vector of the state for the relativistic battery. The follow context will reveal the physical connection between energy transfer and vacuum fluctuations modified by the different motion trajectories.

\section{The extractable work in linear and circular motions}
This work primely aims to study the effects of different motion trajectories on Unruh thermality, from the perspective of quantum work extraction of the UDW battery. With respect to the steady state, the asymptotic status of the battery with a certain amount of quantum coherence is a non-passive state where there exists some non-zero ergotropy. Within the framework of energy transfer, we employ the asymptotic ergotropy for different motion trajectories to detect the thermal nature of the Unruh effect. Furthermore, we will explore how to effectively improve quantum work extraction by the adjustable trajectory.

To investigate the effects of trajectory on the ergotropy, we take into account two typical kinds of the motion trajectories including a uniformly accelerated linear one and uniform-velocity circular one. The trajectory of the linear motion with a uniform acceleration $a$ is described as
\begin{equation}
	\label{eq:(14)}
	x(\tau ) = \left(\frac{1}{a}\sinh (a\tau ),\frac{1}{a}\cosh (a\tau ),0,0 \right).
\end{equation}
The expression of the Wightman function for the linear acceleration motion in Minkowski vacuum is given by
\begin{equation}
	\label{eq:(15)}
    G_0^ + (x,x') =  - \frac{1}{{4{\pi ^2}}}\frac{1}{{{{(t - t' - i\varepsilon )}^2} - {{(x - x')}^2} - (y - y'){}^2 - {{(z - z')}^2}}}.
\end{equation}
Considering the linear motion trajectory in the form of Eq. (14), we can obtain the field correlation function
\begin{equation}
	\label{eq:(16)}
   G_l^ + (x,x') =  - \frac{{{a^2}}}{{16{\pi ^2}}}\frac{1}{{{{\sinh }^2}(\frac{{a\Delta \tau }}{2} - i\varepsilon )}}.
\end{equation}
The Fourier transform of this two-point function is given by
\begin{equation}
	\label{eq:(17)}
   {\mathcal{G}_l}(\Omega) = \frac{\Omega}{2\pi(1 - e^{ - 2\pi \Omega /a}) }.
\end{equation}
From this, the Kossakowski decaying coefficient in the GSKL master equation can be obtained
\begin{align}
	\label{eq:(18)}
	A_l=&\; \frac {\gamma _0}{2}\coth (\frac {\pi \Omega}{a}),\;B_l=\frac {\gamma _0}{2},\;C_l= \mathcal{G}_l(0)-A_l.
\end{align}
Here, ${\gamma _0} = {\lambda ^2}{\Omega}/(2\pi )$ denotes the spontaneous decay rate. In the case of the linear motion trajectory, the asymptotic ergotropy is written as
\begin{equation}
	\label{eq:(19)}
    \xi _l^{(s)} = \frac{{{e^{2\pi \Omega /a}} - 1}}{2({{e^{2\pi \Omega /a}} + 1})}.
\end{equation}
In the later discussion, we will deal with dimensionless parameters by scaling the linear acceleration and proper time as $\tilde{\tau}=\lambda^2 \Omega \tau$ and $\tilde{a}=\frac {a}{\Omega}$. For convenience, we continue to term $\tilde{a}$, $\tilde{\tau}$ as $a$, $\tau$, respectively. We can obtain the function of the steady-state maximum extractable work $\xi _l^{(s)}$ as a function of Unruh temperature $T=\frac {a}{2\pi}$ when the battery is accelerated along the linear motion. The asymptotical behavior of the linear-acceleration ergotropy is plotted as the black solid curve in Figure 1(a). When Unruh temperature $T$ increases, the ergotropy exhibits a monotonic decrease. The global property of Unruh thermality can clearly be demonstrated by the KMS condition.

In contrast, we calculate the maximum extractable work of the battery moving along a circular trajectory with a uniform velocity. The circular trajectory of the battery motion can be expressed as
\begin{equation}
	\label{eq:(20)}
    x(\tau ) = \left(\gamma \tau ,R\cos \frac{{\gamma \tau v}}{R},R\sin \frac{{\gamma \tau v}}{R},0 \right).
\end{equation}
Here, $v$ represents the tangential acceleration in uniform circular motion, $R$ is the radius of the orbit, and $\gamma  = 1/\sqrt {1 - {v^2}}  $ is the Lorentz factor. The centripetal acceleration of the motion is given by $a = {\gamma ^2}{v^2}/R $. In regard to the circular trajectory, we need to employ the expansion ${\sin ^2}[a\Delta \tau /(2v\gamma )] = \frac{{{a^2}{{(\Delta \tau )}^2}}}{{4{v^2}{\gamma ^2}}} - \frac{{{a^4}{{(\Delta \tau )}^4}}}{{48{v^4}{\gamma ^4}}} + \frac{{{a^6}{{(\Delta \tau )}^6}}}{{1440{v^6}{\gamma ^6}}} -  \cdots  $, where $ \Delta \tau  = \tau  - \tau '$. Since it is difficult to find explicit expressions for all orders of $\mathcal{G}(w_0) $ and $ \mathcal{G}(-w_0)$, we consider the approximate analysis in the ultra-relativistic limit,$\gamma  \gg 1 $ \cite{Bell1983} where
\begin{equation}
	\label{eq:(21)}
	G_c^ + (x,x') =  - \frac{1}{{4{\pi ^2}}}\frac{1}{{{{(\Delta \tau  - i\epsilon )}^2}[1 + {a^2}{{(\Delta \tau  - i\epsilon )}^2}/12]}}.
\end{equation}
Therefore, the Fourier transforms of the field correlation function are
\begin{align}
\label{eq:(22)}
   {\mathcal{G}_c}({\Omega}) = \frac{{{\lambda ^2}{\Omega}}}{{2\pi }}(1 + \frac{a}{{4\sqrt 3 {\Omega}}}{e^{ - \frac{{2\sqrt 3 {\Omega}}}{a}}}),\\
   {\mathcal{G}_c}( - {\Omega}) = \frac{{{\lambda ^2}{\Omega}}}{{2\pi }}\frac{a}{{4\sqrt 3 {\Omega}}}{e^{ - \frac{{2\sqrt 3 {\Omega}}}{a}}}.\nonumber
\end{align}
We can derive the decay parameters
\begin{align}
	\label{eq:(23)}
    {A_c} = \frac{{{\Omega}}}{2}(1 + \frac{a}{{2\sqrt 3 {\Omega}}}{e^{ - \frac{{2\sqrt 3 {\Omega}}}{a}}}),
	{B_c} = \frac{{{\Omega}}}{2},
	{C_c} = \mathcal{G}_c(0)-A_c.
\end{align}
The steady-state ergotropy for the circular motion is given by
\begin{equation}
	\label{eq:(24)}
	\xi _c^{(s)} = \frac{{\sqrt 3 {\Omega}}}{{2\sqrt 3 {\Omega} + a{e^{ - 2\sqrt 3 {\Omega}/a}}}}.
\end{equation}
The asymptotical behavior is plotted as the red dashed line in Figure 1(a). Compared two kinds of acceleration trajectories, we find that for low Unruh temperatues, the asymptotic values of the ergotropy in the linear acceleration motion are a litter higher than those in the circular scenario. This results implies that, for small accelerations compared to the energy gap of the battery, linear motion induces the least Unruh noise, where the detection signal of the ergotopy is strong. At a certain temperature, the asymptotic ergotropy for the two trajectories is the same. This feature indicates that one can stimulate the energy transfer along the linear acceleration trajectory by studying the properties of the battery in the circular motion without a boundary. To further interpret this phenomenon, we make use of quantum coherence to uncover the asymptotical behavior of quantum work extraction. By employing the definition of quantum coherence \cite{baumgratz2014quantifying}, the values of the steady-state quantum coherence is written by $C^{(s)}=|\tilde{\gamma}|$ which is determined by the response function of the detector. From Figure 1(b), it is shown that quantum coherence in the circular motion decreases more slowly than that in the linear motion. At large acceleration, more quantum coherence contributes to the improvement of the ergotropy in the case of circular motion. In order to probe higher Unruh temperatures, we can select the circular motion as the most optimal scenario.

\begin{figure}[htb]
	\centering
	\begin{subfigure}{0.45\textwidth}
		\includegraphics[width=\linewidth]{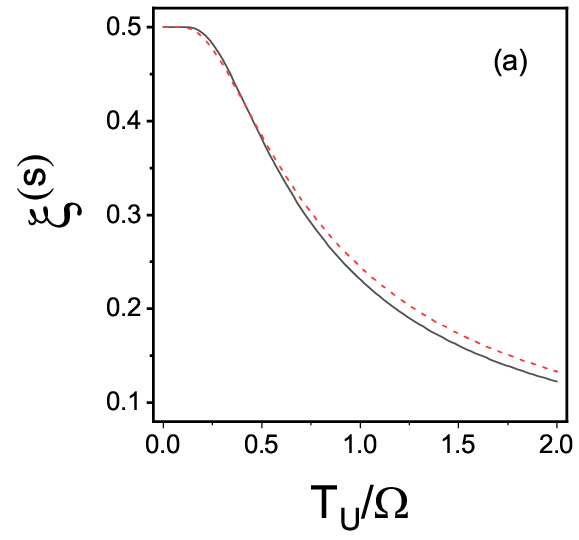}
	\end{subfigure}
	\hfill
	\begin{subfigure}{0.45\textwidth}
		\includegraphics[width=\linewidth]{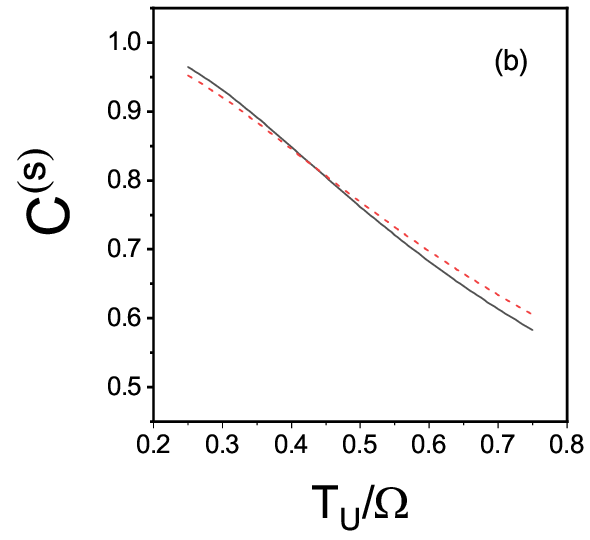}
	\end{subfigure}
	
\caption{The evolution of the ergotropy for the UDW battery in $\mathrm{(A)dS}_4$ sapcetime, which is a function of scaled proper time $\tau$ and acceleration $a$ in the condition of curvature $k=1$. (a) The ergotropy for $\mathrm{dS}_4$. (b) The ergotropy for $\mathrm{AdS}_4$ with the Dirichlet boundary condition $\zeta=1$. (c) The ergotropy for $\mathrm{AdS}_4$ with the transparent boundary condition $\zeta=0$. (d) The ergotropy for $\mathrm{AdS}_4$ with the Neumann boundary condition $\zeta=-1$.}

\end{figure}

The vacuum fluctuations in Minkowski spacetime can be modified owing to the presence of a reflecting boundary. It is necessary to explore the effect of a reflecting boundary on the dynamics of the ergotropy for the accelerated battery along two different trajectories. In the following context, we will study the detection of Unruh thermality in the vicinity of a boundary from the viewpoint of energy transfer.

\section{The effect of a reflecting boundary}
Without loss of generality, we assume that there is a reflecting plane at $z=0$ which is viewed as an ideal boundary. At a certain distance $z$ from the boundary, the battery is confined to move in the $x-y$ plane. Consequently, the Wightman function related to the boundary will be given by means of the method of images,
\begin{equation}
	\label{eq:(25)}
	{G^ + }(x,x') = G_0^ + (x,x') + G_b^ + (x,x').
\end{equation}
Here, $ G_0^ + (x,x')$ represents the two-point correlation function without the boundary, which has already been calculated in Eq. (15), and
\begin{equation}
	\label{eq:(26)}
	G_b^ + (x,x') =  - \frac{1}{{4{\pi ^2}}}\frac{1}{{{{(x - x')}^2} + (y - y'){}^2 + {{(z - z')}^2} - {{(t - t' - i\epsilon )}^2}}},
\end{equation}
provides the correction due to the presence of the boundary. By substituting into the trajectory function of uniform accelerated linear motion, we can obtain the correlation function with regard to the boundary
\begin{equation}
	\label{eq:(27)}
	G_{bl}^ + (x,x') =  - \frac{{{a^2}}}{{16{\pi ^2}}}[\frac{1}{{{{\sinh }^2}(\frac{{a\Delta \tau }}{2} - i\epsilon )}} - \frac{1}{{{{\sinh }^2}(\frac{{a\Delta \tau }}{2} - i\epsilon ) - {a^2}{z^2}}}].
\end{equation}
The Fourier transform of the two-point correlation function is written as
\begin{equation}
	\label{eq:(28)}
	\mathcal{G}_{bl}(\Omega ) = \frac{\Omega }{{2\pi (1-e^{ - 2\pi \Omega  /a} })}\left( 1-\frac{{\sin [\frac{{2\Omega  }}{a}{{\sinh }^{ - 1}}(az)}]}{{2z\Omega  \sqrt {1 + {a^2}{z^2}} }}\right).
\end{equation}
Similar to the no-boundary case, the coefficients for the Kossakowski decay matrix are calculated as
\begin{align}
	\label{eq:(29)}
A_{bl}=&\; \frac {\gamma _{b}}{2}\coth (\frac {\pi \Omega}{a}),\;B_{bl}=\frac {\gamma _{b}}{2},\;C_{bl}= \mathcal{G}_{bl}(0)-A_{bl},
\end{align}	
where $\gamma_{b} = \gamma_0 \{ 1 - \frac{{\sin [\frac{{2{\Omega}}}{a}{{\sinh }^{ - 1}}(az)]}}{{2z{\Omega}\sqrt {1 + {a^2}{z^2}} }}\}$ is the spontaneous decay rate in the condition of the boundary. For the limit of $z\rightarrow 0$, the value $\gamma_{b}$ approaches zero which denotes that the dynamics of the battery is very close to that of a closed system with no dissipation.

It is clearly seen that the KMS condition in the frequency space can be guaranteed by $\mathcal{G}_{bl}(\Omega)=e^{\beta \Omega}\mathcal{G}_{bl}(-\Omega)$ where $\beta=\frac {1}{T}$ represents the Unruh temperature. This fact implies that the effect of the boundary on the Unruh thermality is equivalent to that of free space without a boundary. Notably, the time dependent evolution of the ergotropy for the accelerated battery is influenced by the boundary. Similarly, we will work with the dimensionless quantity by rescaling the distance $\tilde{z}=z\Omega$ and use the simplified term $\tilde{z}=z$ for convenience. Figs. 2(a) and 2(b) respectively show the dynamics of the ergotropy $\xi_{bl}(\tau)$ in the condition of $z=0.5$ and $z=5$ when the battery undergoes the linear motion. It is found that the ergotropy exhibits the significant oscillation in the vicinity of the boundary. This phenomenon results from the protection of quantum coherence in the near boundary region. In the limit of $z\rightarrow 0$, the evolution of the battery will be shielded from the influence of the scalar field as if it were isolated. Moreover, at a far distance, the ergotropy rapidly reaches one steady state, which is gradually close to the asymptotic value in free space. The numerical calculation reveals that the boundary effect is particularly pronounced in the region of low accelerations, which effectively suppresses the decoherence process and significantly enhances the battery performance.

\begin{figure}[htb]
	\centering
	\begin{subfigure}{0.45\textwidth}
		\includegraphics[width=\linewidth]{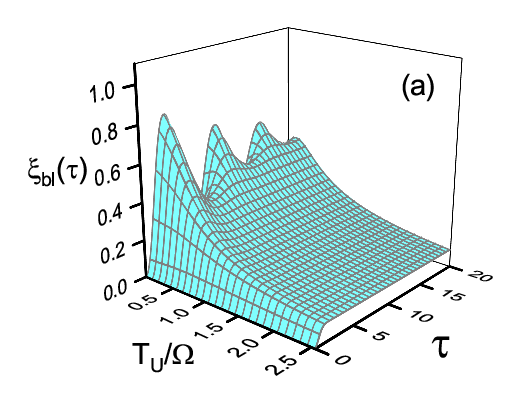}
	\end{subfigure}
	\hfill
	\begin{subfigure}{0.45\textwidth}
		\includegraphics[width=\linewidth]{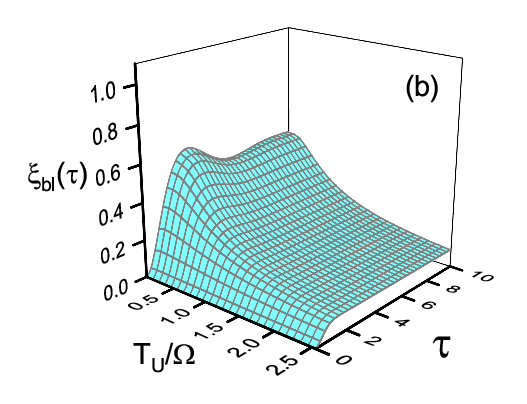}
	\end{subfigure}
	
	\vspace{3mm}
	
	\begin{subfigure}{0.45\textwidth}
		\includegraphics[width=\linewidth]{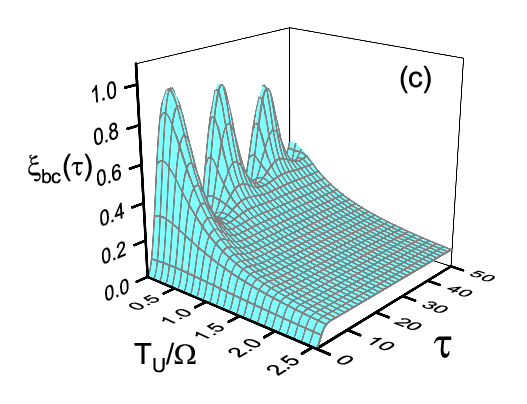}
	\end{subfigure}
	\hfill
	\begin{subfigure}{0.45\textwidth}
		\includegraphics[width=\linewidth]{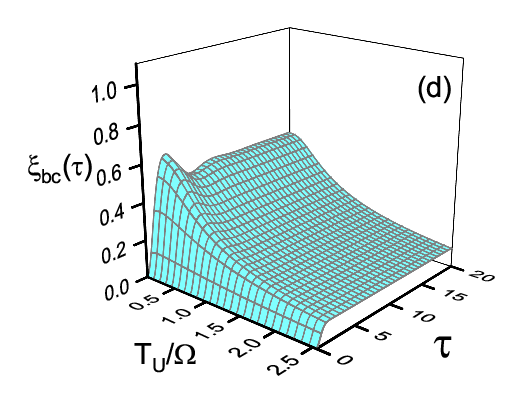}
	\end{subfigure}
	
	\caption{The evolution of the ergotropy for the UDW battery moving along two different trajectories with a boundary, as a function of the scaled proper time $\tau$ and the Unruh temperature $T_U/\Omega$. The parameter $z_0$ describes the distance between the trajectory and boundary plane. (a) $\xi_{bl}(\tau)$ in the linear scenario at a close distance of $z_0=0.5$. (b) $\xi_{bl}(\tau)$ in the linear scenario at the distance $z_0=5$. (c) $\xi_{bc}(\tau)$ in the circular one for $z_0=0.5$. (d) $\xi_{bc}(\tau)$ in the circular one under the condition of $z_0=5$.}
	
\end{figure}

In contrast, we examine the UDW battery undergoing a circular motion within the $x-y$ plane, positioned at a distance $z$ from the boundary. We have
\begin{equation}
	\label{eq:(30)}
	G_{bc}^ + (x,x') =  - \frac{1}{{4{\pi ^2}}}\frac{1}{{4{z^2} - {{(\Delta \tau  - i\epsilon )}^2} - {a^2}{{(\Delta \tau  - i\epsilon )}^4}/12}}.
\end{equation}
Therefore, the Fourier transforms of this Wightman function with respect the effective coupling $\Omega$ are given by
\begin{equation}
	\label{eq:(31)}
		\mathcal{G}_{bc}({\Omega}) = \frac{{{\Omega}}}{{2\pi }}[1 + \frac{a}{{4\sqrt 3 {\Omega}}}{e^{ - \frac{{2\sqrt 3 {\Omega}}}{a}}} -\frac {\sqrt{3}a e^{-\frac {\Omega}{a}f^{+}_1(z)}}{2f^{+}_2(z)\Omega}-\frac {\sqrt{3}a \sin(\frac {\Omega}{a}f^{-}_1(z))}{f^{-}_2(z)\Omega}]
\end{equation}
\begin{equation}
	\label{eq:(32)}
	\mathcal{G}_{bc}( - {\Omega}) = \frac{{{\Omega}}}{{2\pi }}[\frac{a}{{4\sqrt 3 {\Omega}}}{e^{ - \frac{{2\sqrt 3 {\Omega}}}{a}}}  -\frac {\sqrt{3}a e^{-\frac {\Omega}{a}f^{+}_1(z)}}{2f^{+}_2(z)\Omega}].
\end{equation}
where $f^{(\pm)}_1(z)=\sqrt {\pm 6 + 2\sqrt {9 + 12 a^2 z^2}}$ and $f^{(\pm)}_2(z)=\sqrt {(\pm 3 + \sqrt {9 + 12 a^2 z^2})(6 + 8 a^2 z^2)} $ are the symbolized function.
Therefore, we modify the ratio $\tilde{\gamma}^{(c)}$ for the circular trajectory with a boundary as
\begin{equation}
	\label{eq:(33)}
		\tilde{\gamma}^{(c)}=\frac {B_{bc}}{A_{bc}} = \frac  {1-\frac {\sqrt{3}a \sin[\frac {\Omega}{a}f^{-}_1(z)]}{f^{-}_2(z)\Omega}} {1+\frac{a}{{2\sqrt 3 {\Omega}}}{e^{ - \frac{{2\sqrt 3 {\Omega}}}{a}}} -\frac {\sqrt{3}a e^{-\frac {\Omega}{a}f^{+}_1(z)}}{f^{+}_2(z)\Omega}-\frac {\sqrt{3}a \sin(\frac {\Omega}{a}f^{-}_1(z))}{f^{-}_2(z)\Omega}}.
\end{equation}
According to the result of Eq. (13), the asymptotic value of the ergotropy is determined by the ration which is dependent on both the Unruh temperature $T=\frac {a}{2\pi}$ and the distance $z$ from the circular trajectory to the boundary.

Figs. 2(c) and (d) depict the evolution of ergotropy $\xi_{bc}(\tau)$ for uniform circular motion near the boundary at a proximity of $z=0.5$ and far from it at $z=5$, respectively. Compared to the linear trajection, the oscillation of the ergotropy in the vicinity of the boundary becomes more apparent in Figure 2(c). The maximal values of the transient ergotropy for the circular trajectory are larger than those for the linear trajectory at low temperatures. It is seen that the high values of the ergotropy can keep a longer time. This behavior shows that the case of the circular trajectory contributes to the robustness of the ergotropy against the Unruh thermalization. As the distance is increased, the values of the ergotropy will rapidly approach a certain steady one in Figure 2(d). The oscillatory dynamics of the ergotropy is significantly suppressed. At a high temperature, the behavior of the transient ergotropy takes on the monotonic increase in the condition of the long-term driving field.

\begin{figure}[htb]
	\centering
	\adjustimage{width=0.45\textwidth}{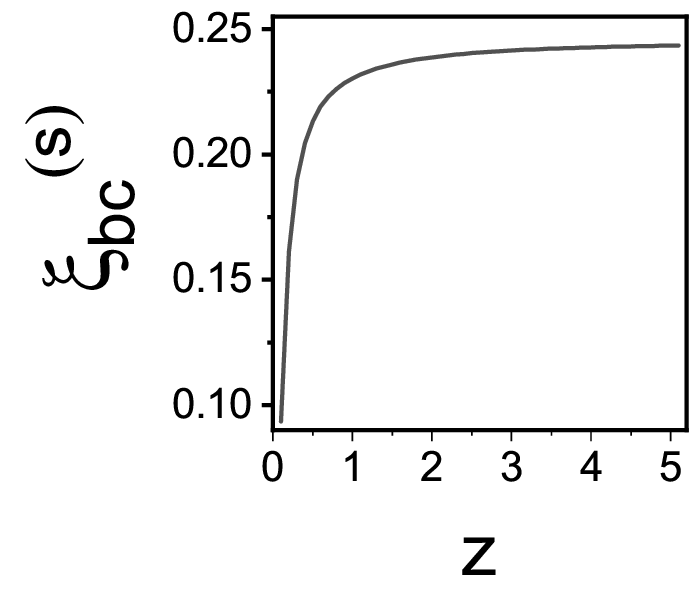}
	\caption{The behavior of the asymptotic ergotropy $\xi_{bc}^{(s)}$ along the circular trajectory with a reflecting boundary as a function of the distance in the case of a certain Unruh temperature $T_U/\Omega=1$.}
\end{figure}

To further investigate the effect of the boundary on the characterization of vacuum fluctuations, we numerically obtain the asymptotic behavior of the ergotropy along the circular trajectory, illustrated in Figure 3. In distinction from the linear trajectory, the steady values of quantum work extraction are gradually improved with increasing the distance and eventually saturate at a certain level. In the limit of $z \rightarrow \infty$, the asymptotical ergotropy is expressed as $\lim_{\tau \rightarrow \infty}\xi_{bc}(\tau)=\frac {1}{2(1+\frac{a}{{2\sqrt 3 {\Omega}}}{e^{ - \frac{{2\sqrt 3 {\Omega}}}{a}}} )}$ which is determined by the Unruh temperature. With the increase of the distance $z$, the denominator of Eq. (33) decays faster than the numerator. In contrast to the linear motion, this fact demonstrates that the asymptotical ergotroy along the circular motion can be enhanced to a certain extent when the distance of the trajectory from the boundary is enlarged. The presence of the saturation value indicates that the effect of the boundary on the battery becomes weaker when the trajectory is farther from the boundary.

\section{Discussion}
In this work, we have explored thermodynamics of quantum work extraction for a relativistic UDW battery coupled to a massless vacuum field in Minkowski spacetime. Several distinct trajectories for the accelerated batter have been considered: a uniform-acceleration linear motion and a constant-velocity circular one. We have investigated the effects of a reflective boundary on the dynamics of the ergotropy in detail. By employing the ergotropy, we have quantitatively estimated the acceleration-induced Unruh thermality modified by different trajectories. Our study is based on the relativistic quantum battery within the framework of open quantum systems.

In linear motion, the monotonic decay of the asymptotic ergotropy can reveal the acceleration-dependent KMS temperature, which is unaffected by the presence of a boundary. Interestingly, at a high temperature ($T>0.5$), the circular motion becomes more favorable for estimating the Unruh effect. This point results from the robustness of quantum coherence against the Unruh thermalization. More quantum non-locality helps improve the energy storage with the larger ergotropy. For a specific temperature $T \approx 0.43$, the two trajectories yield the same ergotropy because the quantum coherence for both motions is the same in the asymptotical condition. The oscillatory behaviors of the ergotropy have occurred across these two trajectories in the presence of a bounary. If the accelerated battery moves close to the boundary, the ergotropy exhibits large instantaneous peaks. This interesting phenomenon arises from the preservation of quantum coherence for the battery near the boundary. Far from the boundary, the oscillations are suppressed, and the values quickly reaches a steady value which is determined by Unruh thermalization. In contrast, the circular motion tends to prolong the oscillatory evolution where the performance of the relativistic is superior. As the distance from the circular trajectory to the boundary plane increases, the steady value of the ergotropy can be increased gradually to a saturation level. This fact demonstrates that the influence of the boundary on the energy storage is weak when the distance of the trajectory from the boundary exceeds a certain threshold.

Finally, our study highlights the profound impact of the different acceleration scenarios on quantum work extraction of the relativistic quantum battery. The distinguishability of the ergoropy in different motion cases underscores the diverse evolutions of the accelerated battery as a quantum probe for vacuum fluctuations. We find that optimizing quantum work extraction for an accelerated battery moving along a circular trajectory in the presence of a boundary plane is advantageous for the detection of the thermal nature of the Unruh effect. A particularly intriguing direction for future research is extending our analysis to curved spacetimes. Investigating quantum work extraction in the presence of gravity could provide deeper insights into interplay between quantum thermodynamics, relativity, and fundamental physics. Comparing these results with those in Minkowski spacetime may offer new paths for detecting relativistic quantum effects in (anti)-de Sitter spacetime and even in black hole spacetime.

\begin{acknowledgments}
We would like to thank Professor Bill Unruh and Professor Philip C. E. Stamp for the discussions. We gratefully acknowledge the financial support from $\mathrm{SCOAP}^{3}$.
\end{acknowledgments}

\end{document}